\documentclass[superscriptaddress,aps,prl,twocolumn,showpacs,preprintnumbers,amsmath,amssymb,floatfix,groupedaddress]{revtex4}
\usepackage{graphics}
\usepackage{graphicx}
\usepackage{amssymb}
\usepackage{bm}

\newcommand{\rmd}{\mathrm{d}}
\newcommand{\mi}{\mathrm{i}}
\begin{document}
\title{Multimode Memories in Atomic Ensembles}
\author{J Nunn}
\email{j.nunn1@physics.ox.ac.uk} \affiliation{Clarendon Laboratory, Oxford University, Parks Road, Oxford OX1 3PU, UK}
\author{K Reim}
\affiliation{Clarendon Laboratory, Oxford University, Parks Road, Oxford OX1 3PU, UK}
\author{K C Lee}
\affiliation{Clarendon Laboratory, Oxford University, Parks Road, Oxford OX1 3PU, UK}
\author{V O Lorenz}
\affiliation{Clarendon Laboratory, Oxford University, Parks Road, Oxford OX1 3PU, UK}
\author{B J Sussman}
\affiliation{Clarendon Laboratory, Oxford University, Parks Road, Oxford OX1 3PU, UK}
\affiliation{National Research Council of Canada, Ottawa, Ontario K1A 0R6, Canada}
\author{I A Walmsley}
\affiliation{Clarendon Laboratory, Oxford University, Parks Road, Oxford OX1 3PU, UK}
\author{D Jaksch}
\affiliation{Clarendon Laboratory, Oxford University, Parks Road, Oxford OX1 3PU, UK}
\begin{abstract}
The ability to store multiple optical modes in a quantum memory allows for increased efficiency of quantum communication and computation. Here we compute the multimode capacity of a variety of quantum memory protocols based on light storage in ensembles of atoms. We find that adding a controlled inhomogeneous broadening improves this capacity significantly.\end{abstract}
 \maketitle

Quantum memories with the ability to store multiple optical modes \cite{Afzelius:2008tg,Vasilyev:2008wm} offer advantages in speed and robustness over single mode memories when incorporated into quantum communication \cite{Sangouard:2008xr,Simon:2007ct,Collins:2007kx} and computation \cite{Tordrup:2008rm,Barrett:2008eu} schemes. For example, dual rail encoding \cite{nielsen2000qca} and photon loss codes \cite{Wasilewski:2007bs} can be integrated into ensemble based computation schemes if multimode storage is available. It is therefore useful to characterize the multimode capacity of a given quantum memory protocol; this is the problem we address in this paper.

Any quantum memory must operate as a linear map, so that quantum superpositions are maintained. Therefore analysis and optimization of the storage efficiency and multimode capacity of any quantum memory can be understood using the well-developed tools of linear algebra.
Of central importance is the singular value decomposition (SVD), or \emph{Schmidt} decomposition \cite{Trefethen:1997sf,Grice:2001nx}, which exists for all linear maps.
 In this paper we introduce a universal method for calculating the multimode capacity of a quantum memory, based on the SVD of the Green's function for the memory interaction. We apply this technique to memories with no inhomogeneous broadening, and we find that their multimode capacity scales poorly with the optical depth of the ensemble. We go on to calculate the multimode capacity of protocols based on controlled reversible inhomogeneous broadening (CRIB) \cite{Sangouard:2007dn,Hetet:2008dp}, and find that their capacities scale more favorably. We then consider a modification to the well-known Raman memory protocol \cite{Kozhekin:2000bs,Nunn:2007wj,Mishina:2008oq,Gorshkov:2007qm}, in which a longitudinal inhomogeneous broadening is applied \cite{Hetet:2008eu,Moiseev:2008kx}; we show that this modified protocol exhibits the same favorable multimode scaling as CRIB. Finally we examine the scaling of the recently proposed atomic frequency comb (AFC) protocol \cite{Afzelius:2008tg}, which was designed with multimode storage in mind; its performance is the best out of the protocols we analyzed.
 
\begin{figure}[h]
\begin{center}
\includegraphics[width = \columnwidth]{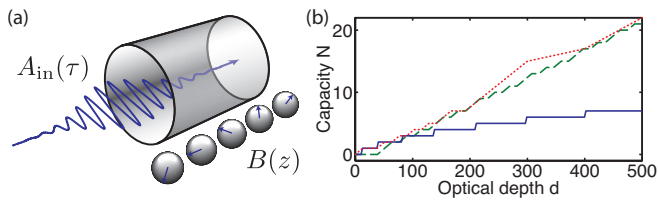}
\caption{(a): An incident signal field is mapped to a stationary spin wave inside an ensemble quantum memory. (b): The multimode scaling of unbroadened ensemble memories (solid line) alongside the tCRIB (dashed line) and lCRIB (dotted line) protocols, with a threshold $\theta=70\%$.}
\label{fig1}
\end{center}
\end{figure}

In the following we deal exclusively with quantum memories based on ensembles of atoms, or atom-like absorbers, although we neglect memories based on measurement and feedback \cite{Hammerer:2008ty,Muschik:2006lp,polzik1,polzik2,polzikreview}. In a typical ensemble quantum memory, an incident signal field with temporal envelope $A_\mathrm{in}(\tau)$ is converted to a stationary excitation known as a \emph{spin wave}, which is distributed as a function of position $z$ within the ensemble with amplitude $B(z)$ (see Figure \ref{fig1} (a)). The quantum memory map takes the form
\begin{equation}
\label{map1}
B(z)=\int_{-\infty}^\infty K(z,\tau)A_\mathrm{in}(\tau)\,\rmd\tau,
\end{equation}
where the kernel $K$ is a Green's function for the storage interaction. In some cases, the form of $K$ can be derived in closed form. The general strategy is to solve the linearized Maxwell-Bloch equations describing the coupling of the signal to the atoms --- the Fourier transform is an invaluable tool in this connection. Even when an analytic solution is not possible, $K$ can always be constructed numerically. In fact, although we use a one dimensional treatment here and in what follows, the same arguments apply if transverse coordinates are included, although the numerical problem becomes significantly larger in this case. Discussions of multimode storage using the transverse spatial degrees of freedom can be found elsewhere \cite{Tordrup:2008rm,Vasilyev:2008wm,Surmacz:2008vf}. Using the SVD, we can always express the Green's function as a convex sum
\begin{equation}
\label{svd}
K(z,\tau)=\sum_k\psi_k(z)\lambda_k\phi^*_k(\tau),
\end{equation}
where the sets of modes $\{\psi_k\}$, $\{\phi_k\}$ each form orthonormal bases, the former for the space of spin wave amplitudes --- the `output space' --- and the latter for the space of signal envelopes --- the `input space'. Note that the SVD does not require the kernel $K$ to be Hermitian: in general, losses can make the kernel complex and non-Hermitian. The positive real singular values $\lambda_k$ couple each input mode $\phi_k$ to the corresponding output mode $\psi_k$. The probability that a single photon in mode $\phi_k$ is stored by the memory is $\lambda_k^2$. The optimal performance of a memory is achieved when the storage and retrieval processes are simply time reverses of each other \cite{Gorshkov:2008rz,gorshkov:033805,Nunn:2007wj}, in which case the retrieval interaction is described by the conjugated Green's function $K^*$. The retrieval efficiency for the $k^\mathrm{th}$ mode is then the same as the storage efficiency $\lambda_k^2$, and so the total efficiency of storage followed by retrieval --- the `memory efficiency' --- is given by $\lambda_k^4$. 

The maximum number of modes of the optical field that can be stored by the memory is just the number of non-negligible singular values. We note that the orthonormal basis $\{\phi_k\}$ is the `natural' input basis for the memory interaction, as determined by its dynamics. It represents the best possible encoding, and the multimode capacity can only be reduced by choosing some other encoding, such as time or frequency bin encoding, for the input modes. To quantify the multimode capacity of a memory, we label the $\lambda_k$ in order of descending magnitude, and define the average memory efficiency for the first $k$ modes, $\Lambda_k=\sum_{j=1}^k\lambda_j^4/k$. Let $\alpha_k$ be equal to $1$ when $\Lambda_k$ exceeds some threshold efficiency $\theta$, with $\alpha_k=0$ otherwise. Then we define the multimode capacity $N=\sum_k \alpha_k$ as the number of modes with average memory efficiency greater than $\theta$. In this paper we use $\theta=70\%$ for all calculations; this choice makes the multimode scaling apparent for experimentally achievable, and numerically tractable, parameters. We have checked that the scaling of $N$ does not depend on the choice made for $\theta$, although imposing higher thresholds reduces $N$, as might be expected.

The SVD of $K$ is related to the spectral decomposition of the normally and anti-normally ordered products (\emph{c.f.} Theorem~5.4 in \cite{Trefethen:1997sf})
\begin{eqnarray}
\nonumber
K_N(\tau,\tau')&=&\int_0^1 K^*(z,\tau)K(z,\tau')\,\rmd z,\\
\label{products}
K_A(z,z')&=&\int_{-\infty}^\infty K(z,\tau)K^*(z',\tau)\,\rmd\tau.
\end{eqnarray}
Direct substitution of Eq.~(\ref{svd}) into the above expressions shows that the eigenvalues of both $K_A$ and $K_N$ are given by $\mu_k=\lambda_k^2$; the capacity $N$ can therefore be directly calculated from these eigenvalues. Here we normalized the longitudinal coordinate so that $z$ runs from $0$ to $1$. In some cases we may have access to an explicit form for the Green's function $K_T$ for the entire memory interaction, that is, storage followed by read out. Suppose that we have
\begin{equation}
\label{entire}
A_\mathrm{out}(\tau)=\int_{-\infty}^\infty K_T(\tau,\tau')A_\mathrm{in}(\tau')\,\rmd\tau',
\end{equation}
where $A_\mathrm{out}$ is the temporal profile of the signal field retrieved from the memory. This is, in some sense, more complete than the storage Green's function in Eq.~(\ref{map1}), since it tells us directly about the connection between the modes that can be stored and the modes that can be retrieved. The SVD can be applied to the bivariate function $K_T$ precisely as it was to $K$ in Eq.~(\ref{svd}), and the singular values $\nu_k$ extracted. In the optimal case that retrieval is the time reverse of storage, we have $K_N=K_T$, and so for consistency we define the multimode capacity in terms of the $\nu_k$ by making the replacement $\lambda_k^2\rightarrow \nu_k$.

Finally, we note that the singular values are invariant under unitary transformations. In particular, we are free to Fourier transform position to spatial frequency ($z\rightarrow k$) or time to angular frequency ($\tau\rightarrow \omega$); the singular values of a kernel $K$ are unchanged by such transformations, and so the multimode capacity can equally well be calculated from the transformed kernel $\widetilde{K}$.

It has previously been shown \cite{Gorshkov:2007qm} that optimal storage in an ensemble memory without inhomogeneous broadening --- whether by using electromagnetically induced transparency (EIT) \cite{EITreview2,Fleischhauer:2002ph,Gorshkov:2007qm} or Raman \cite{Kozhekin:2000bs,Nunn:2007wj,Mishina:2008oq,Gorshkov:2007qm} storage, or simply linear absorption --- is characterized by the kernel
\begin{equation}
\label{unbroadened}
K_A(z,z')=\frac{d}{2}e^{-d(z+z')/2}I_0\left(d\sqrt{zz'}\right),
\end{equation}
where $d$ is the resonant optical depth of the ensemble \cite{Gorshkov:2007qm}, and $I_0$ is a modified Bessel function \footnote{The kernel in Eq.~(\ref{unbroadened}) is derived in the text following Eq.~(23) in \cite{Gorshkov:2007qm} from the anti-normal product of two storage kernels}. Using this result, we calculate the multimode capacity as a function of the optical depth $d$ (see Figure \ref{fig1} (b)). A numerical fit reveals a square-root scaling $N\sim \sqrt{d}/3$. We can identify this as the scaling of the absorption linewidth of an ensemble of $2$-level atoms, each with a Lorentzian lineshape. Since large optical depths can be challenging to achieve in the lab, this scaling makes unbroadened ensemble memories unattractive candidates for multimode storage. Can this poor scaling be improved? Clearly we can always store $N$ modes if we have $N$ separate memories, as long as each has sufficient optical depth $d_\theta$ such that $\lambda_1^4>\theta$. Then the total optical depth is simply $d=Nd_\theta$, so that $N$ scales linearly with $d$. Below we show that in fact separate ensembles are not required for linear scaling. A single physical memory can be made `parallel' by redistributing atoms in frequency. 

We now calculate the multimode capacity for memory protocols based on controlled reversible inhomogeneous broadening (CRIB) \cite{Sangouard:2007dn,Hetet:2008dp}, in which a spatially varying external field is applied to an ensemble in order to artificially broaden the range of frequencies absorbed by the atoms. A signal field is absorbed resonantly by the broadened ensemble. Some time later, the polarity of the external field is reversed, so that the inhomogeneous broadening profile is flipped, with red-detuned atoms becoming blue-detuned and vice versa. The optical dipoles eventually re-phase, and the signal field is re-emitted. The optimal efficiency is achieved if the signal field is re-emitted in the backward direction, but phasematching considerations sometimes favor forward emission. In the following we consider ideal backward emission, since forward emission cannot perform any better. The artificial broadening can be applied in a direction transverse to the propagation direction of the signal field to be stored (tCRIB), or longitudinally --- parallel to the signal beam (lCRIB). As in the rest of this paper, we use a one dimensional propagation model. Since tCRIB is laterally asymmetric, coupling between the transverse spatial structure and the spectrum of the signal is potentially important. Nonetheless the results presented below represent an upper bound on the performance of tCRIB. The Green's function relating the input and output spectra of the signal field is given by \cite{Gorshkov:2007rw,Sangouard:2007dn},
\begin{equation}
\label{tcrib}
\widetilde{K}_T(\omega,\omega')=\frac{1}{2\pi}\frac{e^{-d\left[f(\omega)+f(\omega')\right]}-1}{2\gamma+\mi\left(\omega+\omega'\right)},
\end{equation}
where the lineshape function is $f(\omega)=\int_{-\infty}^\infty \frac{\gamma p(\Delta)}{\gamma + \mi(\Delta+\omega)}\,\rmd\Delta$, with $p(\Delta)\rmd \Delta$ the proportion of atoms whose optical resonance is shifted by a detuning $\Delta$ away from its unbroadened frequency. Here and in what follows $2\gamma$ is the homogeneous linewidth of the unbroadened transition. Note that $d$ is the total optical depth of the ensemble --- if the broadening were removed, the total resonant optical depth of the ensemble would be given by $d$; this definition enables a direct comparison with unbroadened memory protocols. The capacity of tCRIB is plotted in Fig.~(\ref{fig1}) (b), optimized with respect to the spectral width $\Delta_0$ of the applied broadening, assuming a rectangular broadening profile. The scaling is manifestly linear, with $N\sim d/25$. We found that the optimal width also scales linearly, $\Delta_0^\mathrm{opt}/\gamma\sim 9d/5$. These results can be explained qualitatively by considering spectral modes of bandwidth $\gamma$. Every such mode requires a resonant optical depth $d_\theta \gtrsim 10$. To store $N$ modes `side by side' in frequency we should have a broadening $\Delta_0\sim N\gamma$ and a total optical depth $d\sim N d_\theta$, so that $N$ rises linearly with $d$, provided we increase $\Delta_0$ at the same time. Next we consider lCRIB.

The storage kernel for lCRIB, where the atomic resonance frequency varies linearly with the longitudinal coordinate $z$, can be calculated in terms of the spatial frequency $k$ of the spin wave. In this case the spin wave represents the atomic polarization. We obtain \footnote{Details of the derivation will be provided in a forthcoming publication.}
\begin{equation}
\label{lcrib}
\widetilde{K}(k,\tau)=\sqrt{\frac{d\gamma}{2\pi}}e^{-\gamma \tau}k^{-\mi \beta}\left(k+\Delta_0\tau\right)^{\mi \beta-1},
\end{equation}
where $\Delta_0$ is the total spectral width of the broadening applied, and $\beta = d\gamma/\Delta_0$. Note that $k$ is dimensionless, because of the normalization of $z$. Unfortunately this kernel is singular when $k=-\Delta_0\tau$, so a numerical SVD fails. We can, however, directly construct the Green's function in terms of $z$ and $\tau$, which is not divergent, by solving the optical Maxwell-Bloch equations numerically. In Fig.~(\ref{fig1}) (b) we plot the multimode capacity, optimized over the broadening width $\Delta_0$, found from such a numerically constructed Green's function. We used spectral collocation \cite{trefethen2000smm} for the spatial propagation, and a second-order Runge-Kutta method for the time stepping. The capacities of tCRIB and lCRIB are nearly identical, and our numerics suggest that the multimode scaling of both these protocols is the same.

\begin{figure}[h]
\begin{center}
\includegraphics[width=\columnwidth]{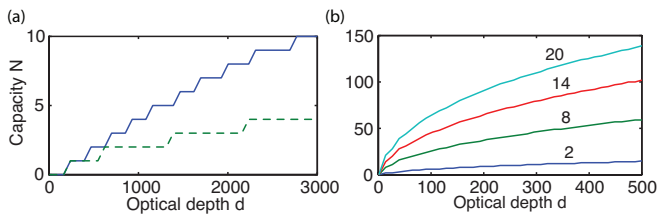}
\caption{(a): The multimode scaling for the Raman protocol with an applied broadening (solid line) and without (dashed line). Here we set $\Omega(\tau)=\sqrt{10d}\gamma e^{-(10\gamma\tau)^2}$ and $\Delta=\sqrt{90d}\gamma$. (b): The capacity of AFC with several different numbers of comb teeth, indicated by the numbers in the plot. We used a threshold $\theta=70\%$.}
\label{fig2}
\end{center}
\end{figure}

Having shown that artificially broadened memories exhibit superior multimode scaling, we now consider a modification to the standard Raman protocol, in which a longitudinal broadening is applied to the storage state. The hope is that such a protocol will combine the advantages of Raman storage, namely broadband, tunable operation, insensitive to unwanted inhomogeneous broadening, with the superior multimode capacity of lCRIB. This type of `hybrid' storage has recently been experimentally demonstrated using magnetic gradients in a warm Rubidium vapor \cite{Hetet:2008eu}. The atoms comprising such a memory have a $\Lambda$-type structure \cite{Hetet:2008eu,Gorshkov:2007qm,Nunn:2007wj}, with an excited state coupled strongly to both a ground and a storage state. A direct transition from the ground to the storage state is forbidden, and therefore storage of an incident signal field is mediated by an off-resonant Raman interaction with the excited state, the strength of which is controlled by the application of an intense control field coupling the storage and excited states. An external field is applied so that the energy of the storage state varies linearly along the length of the ensemble, covering a spectral width $\Delta_0$. The spin wave $B$ in this protocol represents the coherence between ground and storage states. Under adiabatic conditions \cite{Gorshkov:2007qm,Nunn:2007wj}, the following expression approximates the memory map,
\begin{equation}
\label{hybrid}
\widetilde{K}(k,\tau)=\sqrt{\frac{d\gamma}{2\pi}}\Omega^*(\tau)g(k,\tau)e^{\frac{1}{\Gamma}\int_\tau^\infty|\Omega(\tau')|^2[1-d\gamma g(k,\tau')]\,\rmd\tau'},
\end{equation}
where $g(k,\tau)=[d\gamma + \mi \Gamma \times(k+\Delta_0 \tau)]^{-1}$. Here $\Gamma = \gamma + \mi\Delta$ is the complex detuning, with $\Delta$ the common detuning of the signal and control fields from the excited state. $\Omega$ is the slowly varying Rabi frequency of the control field. Figure (\ref{fig2}) (a) shows the multimode capacity for this broadened Raman protocol, operated with a Gaussian control field. Changing the temporal profile of the control pulse, while maintaining a constant energy, affects the shapes of the input modes $\{\phi_k\}$, but not the multimode capacity.  As anticipated, the square-root scaling of the unbroadened memory --- smaller than predicted by Eq.~(\ref{unbroadened}) because we operate far from resonance --- is transformed into the linear scaling of the CRIB protocols by application of a broadening. We found $N\sim d/300$ with $\Delta_0^\mathrm{opt}/\gamma\sim d/77$; in general larger optical depths are required to achieve the same capacity for the Raman protocol, again due to the large detuning from resonance.

Finally we turn to the recently proposed atomic frequency comb (AFC) protocol \cite{Afzelius:2008tg}. This protocol is similar to tCRIB, except that the inhomogeneous profile $p(\Delta)$ takes the form of a comb with spectral width $\Delta_0$, comprised of $M$ equally spaced resonances, each with optical depth $d$: $p(\Delta)=\sum_{j=1}^M \delta(\Delta-\delta_j)$, with $\delta_j=-\Delta_0/2+(j-1)\Delta_0/(M-1)$. The frequency comb is prepared from an initially uniform naturally inhomogeneously broadened ensemble by optical pumping, which essentially removes atoms that would absorb at frequencies between the comb teeth. More teeth can be added by pumping out fewer atoms, which does not require any change in the size or density of the ensemble. Therefore the important physical resource is the depth $d$ associated with \emph{each} tooth, since this is set by the ensemble density. Note, however, that the total optical depth $d_\mathrm{tot}=Md$, found by summing the contributions from all the teeth, increases as we add more teeth. This explains the remarkable multimode scaling discussed below. The inhomogeneous broadening does not need to be `flipped' before retrieval, as it does for CRIB, since its discrete structure leads to periodic rephasing of the atomic dipoles. The Green's function for storage, followed by backward retrieval, using the same notation as in Eq.~(\ref{tcrib}), is \cite{Gorshkov:2007rw,Afzelius:2008tg}
\begin{equation}
\label{AFC}
\widetilde{K}_T(\omega,\omega')=\frac{1}{2\pi}\frac{f(\omega)-f(\omega')}{\omega-\omega'}\times\frac{e^{-d\left[f(\omega)+f(\omega')\right]}-1}{f(\omega)+f(\omega')}.
\end{equation}
In Fig.~(\ref{fig2}) (b) we plot the multimode capacity of AFC for several values of $M$. For every $M$ the capacity exhibits the square-root scaling characteristic of unbroadened memories, but $N\propto M$, and can be made arbitrarily large simply by adding more teeth to the comb. This increases $d_\mathrm{tot}$, but requires no increase in the ensemble density. Provided $d>d_\theta \sim 10$ the multimode capacity is, in principle, infinite. Of course, as with all of these protocols, $N$ can never exceed the total number of atoms in the ensemble. More importantly, however, as more teeth are added the comb width $\Delta_0$ must be increased to ensure the teeth are well-separated; we found a \emph{finesse} \cite{Afzelius:2008tg} $F=\Delta_0/2\gamma(M-1)\gtrsim 30$ is required. $N$ is therefore limited because $\Delta_0$ is bounded by the spectral width of the initial inhomogeneous line.

In conclusion, we have developed a universal approach to quantifying the ability of ensemble quantum memories to store multiple spectral/temporal modes of an optical field. We showed that the multimode capacity of standard EIT and Raman memory protocols scales with $\sqrt{d}$, while the capacity of CRIB protocols scales linearly with $d$. We considered the capacity of a modified Raman protocol in which a longitudinal broadening is applied to the storage state, and found that this linear scaling is reproduced. Finally we considered the AFC protocol; its capacity is not limited by $d$, provided a certain threshold is exceeded.

\acknowledgements
This work was supported by the EPSRC through the QIP IRC
(GR/S82716/01) and project EP/C51933/01. JN
thanks Hewlett-Packard. IAW was
supported in part by the European Commission under the Integrated
Project Qubit Applications (QAP) funded by the IST directorate as
Contract Number 015848, and the Royal Society.

\end{document}